\documentclass[journal,onecolumn]{IEEEtran}
\usepackage{epsfig}
\usepackage{graphicx}
\usepackage{graphics}
\usepackage{epstopdf} 
\usepackage{tikz}
\usetikzlibrary{calc,shapes.geometric}
\usetikzlibrary{arrows,snakes,backgrounds}

\ifCLASSINFOpdf
\else
\fi

\usepackage[cmex10]{amsmath}

\def\clock{\n=\time \divide\n 60
  \m=-\n \multiply\m 60 \advance\m \time
  \ifnum \n>12 \advance\n -12 \fi
   \number\n.\twodigits\m~\ampm\time}
\def\ampm#1{\ifnum #1< 720 am\else pm\fi}
\def\twodigits#1{\ifnum #1<10 0\fi \number#1}

\usepackage{epsfig}

\def\hyptest{\renewcommand{\arraystretch}{-0.7} 
\begin{array}{c}  
\mbox{\tiny{$H_{1}$}}  \\ \vspace{-0.5 mm}
>\\ 
<\\  
\mbox{\tiny{$H_{0}$}} 
\end{array}
}

\def\nexto{\kern -0.54em}

\def\prob{{\rm {I\ \nexto P}}}
\def\dist{\stackrel{d}{=}}

\def\pfa{{\rm P_{FA}}}

\newcount\m \newcount\n
\def\clock{\n=\time \divide\n 60
  \m=-\n \multiply\m 60 \advance\m \time
  \ifnum \n>12 \advance\n -12 \fi
   \number\n.\twodigits\m~\ampm\time}
\def\ampm#1{\ifnum #1< 720 am\else pm\fi}
\def\twodigits#1{\ifnum #1<10 0\fi \number#1}

\begin{document}

%TITLE AND AUTHOR
\title{Extension of the Geometric Mean Constant False Alarm Rate Detector to Multiple Pulses}
\author{Graham  V. Weinberg  \\ (Draft created at \clock)\\
%Graham.Weinberg@dsto.defence.gov.au
 }
\maketitle

% The paper headers
\markboth{GM-CFAR with Multiple Pulses \today}%
{}

\begin{abstract}
The development of sliding window detection processes, based upon a single cell under test, and operating in clutter modelled by a Pareto distribution, has been examined extensively. This includes the construction of decision rules with the complete constant false alarm rate property. However, the case where there are multiple pulses available has only been examined in the partial constant false alarm rate scenario. This paper outlines in the latter case how the probability of false alarm can be produced, for a geometric mean detector, using properties of gamma distributions. The extension of this result, to the full constant false alarm rate detector case, is then presented.
\end{abstract}

\begin{IEEEkeywords}
Radar detection; Sliding window detector;  Geometric mean detector; Constant false alarm rate; Multiple pulses
\end{IEEEkeywords}

\section{Introduction}
This paper is concerned with the extension of the work of \cite{weinberg13} to the multiple pulse scenario, as initiated in \cite{mezache}.  A key difference is that a novel technique is introduced, utilising the fact that the cumulative distribution function of a gamma distributed random variable can be expressed as a sum of Poisson-like terms, allowing for simple derivation of the appropriate probability of false alarm (Pfa) for the geometric mean (GM) sliding window detector.
In addition to this, the work in \cite{mezache} is extended to produce a GM detector with the full constant false alarm rate (CFAR) property, in the clutter environment of interest.

To contextualise this work the basic theory of sliding window detectors is introduced. Useful references on this include \cite{gandhi} - \cite{weinbergbook}.
Sliding window detectors assume the existence of a series of non-negative clutter measurements, denoted $Z_1, Z_2, \ldots, Z_N$, which are assumed to be independent and identically distributed. These are referred to as the constituents of the clutter range profile (CRP). The context for this work is X-band maritime surveillance radar, and so it will be assumed that these have a Pareto Type I distribution. Hence for all $j \in \{1, 2, \ldots, N\}$, 
\begin{equation}
F_{Z_j}(t) = \prob(Z_j \leq t) = 1 - \left( \frac{\beta}{t}\right)^\alpha, \label{parcdf}
\end{equation}
for $t\geq \beta$, and is zero otherwise. Here $\alpha>0$ is the shape and $\beta > 0$ is the scale parameter. This distribution function has a support not beginning at zero. The justification of a Pareto Type I model, for the scenario of interest, has been documented in \cite{weinbergbook}. To summarise the latter, the original fits to real X-band maritime surveillance radar clutter showed that a Pareto Type II model is appropriate. Since in many cases the Pareto scale parameter $\beta << 1$ it follows that the Pareto Type I model can be used as a basis for detector design.

Next a cell under test is taken, which is assumed to be independent of the CRP, and denoted by $Z_0$. This is also a non-negative random variable. 
Sliding window detectors apply some function $f = f(Z_1, Z_2, \ldots, Z_N)$ to the CRP to produce a single measurement of the clutter level.
This is then normalised by a constant $\tau > 0$, called the threshold multiplier. Suppose that $H_0$ is the hypothesis that the CUT does not contain a target, while $H_1$ is the hypothesis that it contains a target embedded in clutter.
A typical test can be written
\begin{equation}
Z_0 \hyptest \tau f(Z_1, Z_2, \ldots, Z_N), \label{test1}
\end{equation}
where the notation in the above means that $H_0$ is rejected only if $Z_0 > \tau f(Z_1, Z_2, \ldots, Z_N)$. 

The Pfa of test \eqref{test1} is given by
\begin{equation}
\pfa = \prob(Z_0 > \tau f(Z_1, Z_2, \ldots, Z_N) | H_0). \label{pfa1}
\end{equation}
For a given Pfa and function $f$, one can solve for $\tau$ for application in \eqref{test1}. If this can be done in such a way that the Pfa does not vary with the clutter power, then the test is said to have the CFAR property. The importance of this property is evident from the fact that if there is variation with the resulting Pfa, this can cause series problems when the detector outputs are applied to a tracking algorithm. Hence sliding window detectors, with the CFAR property, are highly desirable in practical detector design. 

Detectors of the form \eqref{test1} do not have the full CFAR property in the Pareto case, as can be observed in \cite{weinbergMSSP}. 
In order to address this, a transformation approach for the design of detectors was introduced in \cite{weinberg13}, which was generalised in \cite{weinberg14}. The main detector to be considered in the current work takes the form
\begin{equation}
Z_0 \hyptest \beta^{1-N\tau} \prod_{j=1}^N Z_j^\tau, \label{gmdet}
\end{equation}
which when applied to the Pareto Type I case results in the Pfa given by
\begin{equation}
\pfa = \frac{1}{(1+\tau)^N}. \label{pfagm1}
\end{equation}
This will be referred to as the GM detector throughout.
In view of \eqref{pfagm1}, it follows that the detector \eqref{gmdet} is CFAR with respect to the Pareto shape parameter, but requires {\em a priori} knowledge of the Pareto scale parameter.

Further analysis revealed that the CRP minimum is a complete sufficient statistic for $\beta$, and consequently one can consider the alternative detector
\begin{equation}
Z_0 \hyptest Z_{(1)}^{1-N\tau} \prod_{j=1}^N Z_j^\tau, \label{gmdet2}
\end{equation}
where $Z_{(1)} = \min\{Z_1, Z_2, \ldots, Z_N\}$. Then it is shown in \cite{weinberg17} that \eqref{gmdet2} has Pfa
\begin{equation}
\pfa = \frac{N}{N+1} \frac{1}{(1+\tau)^N}, \label{pfagm2}
\end{equation}
proving that the detector \eqref{gmdet2} is completely CFAR.

The main idea in \cite{mezache} is to extend \eqref{gmdet} to allow for multiple pulses, or equivalently multiple CUTs. In the next section this detector is specified and its Pfa produced using some properties of gamma distributed random variables. The approach is then used to produce a variant of \eqref{gmdet2} in the multiple pulse scenario, with the full CFAR property.

\section{Case 1: Partial CFAR Detector}
The multiple pulse based detector assumes that there are a series of $N$ CUTs available, which will be denoted $X_1, X_2, \ldots X_N$.
The first version to be considered takes the form
\begin{equation}
\prod_{i=1}^N X_i \hyptest \beta^{N-M\tau} \prod_{j=1}^M Z_j^\tau, \label{gmdet3}
\end{equation}
where the CUT variables are assumed to be independent and identically distributed under $H_0$, $N$ is a positive natural number, and the CUT statistics are assumed to be independent of the CRP. In \cite{mezache} a more general form is taken, where multiple CRPs are assumed. The analysis below extends easily to this setting, and is hence omitted for brevity.

In order to derive the Pfa of \eqref{gmdet3}, introduce random variables $X_i^*$ and $Z_j^*$ which are the Pareto duals. These are exponentially distributed random variables with the property that $X_i = \beta e^{\alpha^{-1} X_i^*}$ (under $H_0$) and  $Z_j = \beta e^{\alpha^{-1} Z_j^*}$, and so generate the relevant Pareto variables. Then it can be shown that
\begin{equation}
\pfa = \prob\left( \sum_{i=1}^N X_i^* > \tau \sum_{j=1}^M Y_j^*\right). \label{pfaexp1}
\end{equation}
Since the duals in \eqref{pfaexp1} have exponential distributions with parameter unity, introduce random variables $W_1$ and $W_2$ which have gamma distributions: $W_1 \dist \gamma(N, 1)$ and $W_2 \dist \gamma(M, 1)$.
Then the Pfa of \eqref{pfaexp1} can be written
\begin{equation}
\pfa = \int_0^\infty f_{W_2}(t) \prob\left( W_1 > t\tau\right)dt, \label{pfaexp2}
\end{equation}
where $f_{W_2}$ is the density of $W_2$. It can be shown that since $N$ is a natural number, the distribution function of the gamma variable in \eqref{pfaexp2} can be written in the form
\begin{equation}
 \prob\left( W_1 > t\tau\right) = \sum_{l=0}^{N-1} \frac{1}{l!} w^l e^{-l}, \label{pfaexp3}
\end{equation}
which is a sum of Poisson point probabilities (see \cite{cinlar}).
Hence, since 
\begin{equation}
f_{W_2}(t) = \frac{1}{(M-1)!} t^{M-1} e^{-t} \label{gamdens}
\end{equation}
it follows by applying \eqref{pfaexp3} and \eqref{gamdens} to \eqref{pfaexp2}, the Pfa reduces to 
\begin{equation}
\pfa = \sum_{l=0}^{N-1} { M + l-1 \choose l} \frac{ \tau^l}{ (\tau+1)^{M+l}}. \label{pfaexp4}
\end{equation} 
This expression is consistent with the corresponding result in \cite{mezache}. In the next section it is shown how a similar line of analysis can be used to produce a multiple pulse version of \eqref{gmdet2} with the full CFAR property.

\section{Case 2: Full CFAR Detector}
Replacing $\beta$ with the CRP minimum, one arrives at the detector
\begin{equation}
\prod_{i=1}^N X_i \hyptest {Z_{(1)}}^{N-M\tau} \prod_{j=1}^M Z_j^\tau. \label{gmdet4}
\end{equation}
In order to derive the Pfa of \eqref{gmdet4}, one applies the Pareto duals as before, which yields 
\begin{equation}
\pfa = \prob\left( \sum_{i=1}^N X_i^* > (N-M\tau) Y_{(1)}^*\tau + \sum_{j=1}^M Y_j^*\right), \label{pfaexp5}
\end{equation}
where $Y_{(1)}^*$ is the minimum of the duals $Y_j^*$, and so has an exponential distribution with parameter $M$.
By conditioning on this minimum, the Pfa becomes

\begin{equation}
\pfa = \int_0^\infty  f_{ Y_{(1)}^*}(t) \prob\left(\sum_{i=1}^N X_i^* > Nt + \tau \sum_{j=1}^M \left( Y_j^* - t\right) | Y_{(1)}^* = t \right)dt. \label{pfaexp6}
\end{equation}
Let $W_1 = \sum_{i=1}^N Z_{i}^* \dist \gamma(N,1)$ and define $W_2 = \sum_{j=1}^M \left( Y_j^* - t\right) | Y_{(1)}^* = t$.
In \cite{weinberg17} it is shown that $W_2 \dist \gamma(M-1, 1)$. Hence it follows that
\begin{eqnarray}
\pfa &=& \int_0^\infty Me^{-Mt}\prob( W_1 > Nt + \tau W_2) dt\nonumber\\
\nonumber\\
&=& \int_0^\infty \int_0^\infty M e^{-Mt} f_{W_2}(w) \prob(W_1 > Nt + \tau w)dt dw \nonumber\\
\nonumber\\
&=& \int_0^\infty \int_0^\infty M e^{-Mt} \frac{1}{(M-1)!} w^{M-1} e^{-w} \sum_{l=0}^{N-1} \frac{1}{l!} [Nt+\tau w]^l  e^{-Nt-\tau w} dt dw,
 \label{pfaexp7}
\end{eqnarray}
where the appropriate densities have been applied, in addition to \eqref{pfaexp3}.
By applying the binomial expansion 
\begin{equation}
[Nt+\tau w]^l = \sum_{n=0}^l {l \choose n} (Nt)^{l-n} (\tau w)^n
\end{equation}
it can be shown that the Pfa reduces to 
\begin{equation}
\pfa = \sum_{l=0}^{N-1} \frac{M}{(M-1)!} \frac{1}{l!} \sum_{n=0}^l {l \choose n} \tau^n \int_0^\infty t^{l-n} e^{-[N+M]t} dt
\int_0^\infty w^{M+n-1} e^{-[\tau + 1]}dw. \label{pfaexp8}
\end{equation}
Finally, by evaluating the gamma function integrals, the Pfa can be shown to reduce to 
\begin{equation}
\pfa = M \sum_{l=0}^{N-1} \sum_{n=0}^l {M + n-1 \choose n} [N+M]^{-(l-n+1)} \frac{\tau^n}{ [\tau+1]^{M+n}}. \label{finalpfa}
\end{equation}
Thus \eqref{finalpfa} shows that \eqref{gmdet4} is a CFAR decision rule, for application to the context of \cite{mezache}. 

It is worth observing that in the scenario where $N=1$, which requires $l=0$ and $n=0$, the Pfa in \eqref{finalpfa}
reduces to an expression analogous to \eqref{pfagm2}, showing that the multiple pulse detector's Pfa is consistent with the single 
pulse case\footnote{Note that $N$ appears on the left hand side of \eqref{gmdet4}, while $M$ is used on the right hand side.}.

\section{Conclusions and Further Work}
The main contribution of this paper was to show how the general result from \cite{mezache}, for the GM detector \eqref{gmdet} as used in \eqref{gmdet3}, could be extended to produce a GM-CFAR for multiple pulses in Pareto distributed clutter. The next stage of this work will examine tangible examples of performance, including the effects of range-spread interference and clutter power transitions.

\end{document}